\input{aipcheck}

\documentclass[
    ,final            
  ]
  {aipproc}
\usepackage{aas_macros}
\layoutstyle{6x9}

\newcommand{\lsim}{\mathrel{\hbox{\rlap{\lower.55ex
\hbox{$\sim$}} \kern-.3em \raise.4ex \hbox{$<$}}}}

\begin{document}

\title{$^{30}$S($\alpha$,p) in X-Ray Bursts at CRIB}

\author{D. Kahl}{
  address={Center for Nuclear Study (CNS), the University of Tokyo, Wak\={o}, Saitama, 351-0198 Japan}
}

\author{A. A. Chen}{
  address={Department of Physics \& Astronomy, McMaster University, Hamilton, Ontario, L8S-4M1 Canada}
}

\author{S. Kubono}{
  address={Center for Nuclear Study (CNS), the University of Tokyo, Wak\={o}, Saitama, 351-0198 Japan}
}

\author{D. N. Binh}{
  address={Center for Nuclear Study (CNS), the University of Tokyo, Wak\={o}, Saitama, 351-0198 Japan}
}

\author{J. Chen}{
  address={Department of Physics \& Astronomy, McMaster University, Hamilton, Ontario, L8S-4M1 Canada}
}

\author{T. Hashimoto}{
  address={Center for Nuclear Study (CNS), the University of Tokyo, Wak\={o}, Saitama, 351-0198 Japan}
}

\author{S. Hayakawa}{
  address={Center for Nuclear Study (CNS), the University of Tokyo, Wak\={o}, Saitama, 351-0198 Japan}
}

\author{D. Kaji}{
  address={RIKEN (the Institute of Physical and Chemical Research), Wak\={o}, Saitama, 351-0198 Japan}
}

\author{A. Kim}{
  address={Department of Physics, Ewha Womans University, Seoul 120-750 Korea}
}

\author{Y. Kurihara}{
  address={Center for Nuclear Study (CNS), the University of Tokyo, Wak\={o}, Saitama, 351-0198 Japan}
}

\author{N. H. Lee}{
  address={Department of Physics, Ewha Womans University, Seoul 120-750 Korea}
}

\author{S. Nishimura}{
  address={RIKEN (the Institute of Physical and Chemical Research), Wak\={o}, Saitama, 351-0198 Japan}
}

\author{Y. Ohshiro}{
  address={Center for Nuclear Study (CNS), the University of Tokyo, Wak\={o}, Saitama, 351-0198 Japan}
}

\author{K. Setoodeh nia}{
  address={Department of Physics \& Astronomy, McMaster University, Hamilton, Ontario, L8S-4M1 Canada}
}

\author{Y. Wakabayashi}{
  address={Center for Nuclear Study (CNS), the University of Tokyo, Wak\={o}, Saitama, 351-0198 Japan}
  ,altaddress={Advanced Science Research Center, Japan Atomic Energy Agency (JAEA), Naka-gun, Ibaraki 319-1195, Japan}
}

\author{H. Yamaguchi}{
  address={Center for Nuclear Study (CNS), the University of Tokyo, Wak\={o}, Saitama, 351-0198 Japan}
}

\begin{abstract}
Over the past three years, we have worked on developing a well-characterized $^{30}$S radioactive beam to be used in a future experiment aiming to directly measure the $^{30}$S($\alpha$,p) stellar reaction rate within the Gamow window of Type I X-ray bursts.  
\end{abstract}

\maketitle

The $^{30}$S($\alpha$,p) reaction is a significant link in the {\it $\alpha$p}-process, which competes with the {\it rp}-process in Type I X-ray bursts (XRBs) \cite{1981ApJS...45..389W}, but the reaction rate is virtually unconstrained by experimental data.
The $^{30}$S($\alpha$,p) reaction rate is important to the overall energy generation of X-Ray Bursts \cite{2008ApJS..178..110P}, influences the neutron star crustal composition \cite{2006NuPhA.777..601S}, and may explain the bolometric double-peaked nature of some rare X-Ray Bursts \cite{2004ApJ...608L..61F}.
The theoretical $^{30}$S($\alpha$,p) cross section at astrophysical energies is typically calculated using a statistical model, but this approach is unfavorable if there are significant narrow-resonant contributions.
Previous work indicates that for $\alpha$-induced reactions on $T_{z}=\pm1$ ($T_{z}\equiv (N-Z)/2$) nuclei with $18 \leq A \leq 30$, the cross sections are shown to be dominated by natural-parity, $\alpha$-cluster resonances \cite{2005PrPNP..54..535A}, thus experimental measurements to constrain the $^{30}$S($\alpha$,p) reaction rate are warranted.

\section{$^{30}$S Beam Production}
The low-energy Center for Nuclear Study (CNS) radioactive ion beam (CRIB) separator facility of the University of Tokyo \cite{2005NIMPA.539...74Y} and located at the Nishina Center of RIKEN is capable of producing a $^{30}$S RI beam suitable for studying the astrophysical $^{30}$S($\alpha$,p) reaction.
We produce $^{30}$S via the $^{3}$He($^{28}$Si,$^{30}$S,)n reaction.
The target $^{3}$He gas is confined by 2.5 $\mu$m Havar windows and cooled with LN$_{2}$ to an effective temperature around 80--90 K \cite{2008NIMPA.589..150Y}.
The cocktail beam emerging from the production target is mainly characterized and purified in the experiment hall by two magnetic dipoles and a Wien (velocity) filter, with beam-focusing magnetic multipoles surrounding these elements.
These beam-line components are separated by four focal planes of interest.
The primary beam focal point and the production target are located at F0, the dispersive focal plane between the two magnetic dipoles is denoted `F1,' the achromatic focal point after the second dipole `F2,' and the location of the experiment scattering chamber after the Wien filter `F3.'

As we conducted $^{30}$S RI beam development tests in December 2006, May 2008, and July 2009 (two days each) varying many parameters to optimize results for $^{30}$S, we will limit the discussion to the most noteworthy points.
We tested three primary beams: $^{28}$Si$^{9+}$ of 6.9 MeV/u at 100 pnA, $^{28}$Si$^{10+}$ of 7.54 MeV/u at 10 pnA, and $^{28}$Si$^{9+}$ of 7.4 MeV/u at 144 pnA, listed in chronological test order\footnote{All intensities quoted here were the maximum available at the time of the various tests.}.
We found that $^{30}$S beam intensity shows a positive correlation with primary beam energy within this range, justifying our choice of the highest $^{28}$Si beam energy available from the cyclotron for each test.
The production target thickness was also optimized for $^{30}$S yield, which was $\sim$1.7 mg/cm$^{2}$ of $^{3}$He, corresponding to a cryogenic gas pressure of 400 Torr.

\begin{figure}
  \includegraphics[height=.4\textheight, angle=270]{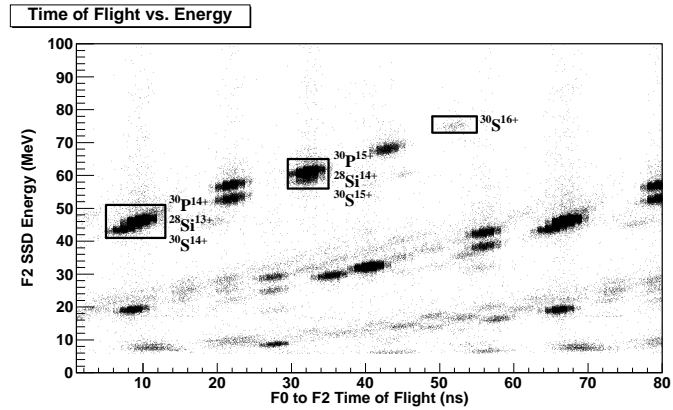}
\caption{The plot shows the particle flight time in nanoseconds on the abscissa and the residual energy in million electron volts on the ordinate for various nuclear species in the cocktail beam at the achromatic focal plane F2.  Although $^{30}$S$^{16+}$ is clearly separated, the loci of other charge-states of $^{30}$S are heavily contaminated.  The dispersive momentum slits are set such that $\Delta p/p \leq 0.625\%$.}
\label{pid_plot}
\end{figure}
Only the fully-stripped ion $^{30}$S$^{16+}$ ($A/q\sim1.875$)\footnote{For simplicity, we quote $A/q$ without units.} is clearly separated from the leaky primary beam (see Figure \ref{pid_plot}), which can never have $A/q<2$.
$^{30}$P also shows up as an impurity for all but the fully-stripped case of $^{30}$S.
In December 2006, we determined that without measuring the energy or a significant energy-loss of the cocktail beam, it was impossible to fully separate $^{30}$S$^{15+}$ from $^{28}$Si$^{14+}$ at a satisfactory level to avoid false-positives.
In July 2009 we could not purify $^{30}$S$^{+14}$ above the $\sim2\%$ level even with use of the Wien filter.
Our results for $^{30}$S$^{16+}$ have continued to improve each year, and we have successfully achieved $\sim10^{4}$ particle Hz on target with $\sim25-30\%$ purity with $E_{beam}=30\pm3$ MeV.
We considered ways of increasing the intensity of fully-stripped charge states of ions emerging from the production target Havar exit window.
We measured the charge-state distribution of $^{28}$Si beam ions in a thick carbon foil (550 $\mu$g/cm$^{2}$) compared to Havar foil (2.2 $\mu$m) (Table~\ref{tab1}).
For a $^{28}$Si beam of 3.4 MeV/u ($\sim E_{beam}$ of $^{30}$S), it was found that transmission of highly charged states of $^{28}$Si is improved through carbon foil compared to Havar foil with a ratio consistent with predictions of LISE++ \cite{2008NIMPB.266.4657T}.
In our July 2009 test, we used a 2.5 $\mu$m Be foil after the production target, which when normalized for comparison with the May 2008 results, indicates an increase in the $^{30}$S$^{16+}$ intensity by a factor of 2.
Although one theoretically expects this intensity increase to be on the order of a factor of 10--20, the Be foil was partially broken, possibly accounting for this deficiency.
\begin{table}[h]
\centering
  \begin{tabular}{ c  c  c  }
      \hline
        {\normalsize Target} & {\normalsize Species} & {\normalsize Normalized} \\ 
	  & & pps @ 10 enA \\ \hline
	Havar & $^{28}$Si$^{12+}$ & $1.075 \times 10^{8}$ \\ 
	Havar & $^{28}$Si$^{13+}$ & $6.013 \times 10^{7}$ \\ 
	Havar & $^{28}$Si$^{14+}$ & $3.901 \times 10^{6}$ \\ 
	Carbon & $^{28}$Si$^{12+}$ & $1.758 \times 10^{8}$ \\ 
	Carbon & $^{28}$Si$^{13+}$ & $1.300 \times 10^{8}$ \\ 
	Carbon & $^{28}$Si$^{14+}$ & $4.365 \times 10^{7}$ \\ \hline
\end{tabular}
\caption
{Intensity of selected charge states of $^{28}$Si after passing through Havar foil or carbon foil.} 
\label{tab1}
\end{table}
\section{Summary and Future Outlook}
We successfully developed a $^{30}$S RI beam of 10$^{4}$ particle Hz of $\sim$25\% purity and $E_{beam}=30\pm 3$ MeV.
In September 2010, we will measure the $^{4}$He($^{30}$S,p) cross section on an event-by-event basis using an active target method using the thick-target method in inverse-kinematics \cite{1990SvJNP..52..408}.  
These experiments were made possible through the CNS and RIKEN collaboration.  The McMaster group is appreciative of funding from the National Science and Engineering Research Council of Canada.  The authors sincerely thank the Nishina Center beam operators.

\bibliographystyle{aipproc}   

\bibliography{/home/daid/library/library}

\IfFileExists{\jobname.bbl}{}
 {\typeout{}
  \typeout{******************************************}
  \typeout{** Please run "bibtex \jobname" to optain}
  \typeout{** the bibliography and then re-run LaTeX}
  \typeout{** twice to fix the references!}
  \typeout{******************************************}
  \typeout{}
 }

\end{document}